\documentclass[12pt]{article}

\usepackage{amsmath,amssymb,setspace}

\def\be{\begin{equation}}
\def\ee{\end{equation}}
\def\beq{\begin{eqnarray}}
\def\eeq{\end{eqnarray}}
\def\lb{\label}
\def\d{\mathrm{d}}

\def\tauf{\tau_{\rm fluid}}

\def\geff{\gamma_{\rm eff}}
\def\gp{\gamma_\perp}

\begin{document}
\begin{spacing}{1}

\begin{center}
{\large\bf
Tilt and phantom cosmology}

\

A.A. Coley, S. Hervik, W.C. Lim 

Department of Mathematics \& Statistics, Dalhousie University,
Halifax, Nova Scotia, Canada B3H 3J5

Email:
aac@mathstat.dal.ca, herviks@mathstat.dal.ca, wclim@mathstat.dal.ca


\end{center}

\begin{abstract}

We show that in tilting perfect fluid cosmological models with an
ultra-radiative equation of state, generically the tilt becomes
extreme at late times and, as the tilt instability sets in,
observers moving with the tilting fluid will experience singular
behaviour in which infinite expansion is reached within a finite
proper time, similar to that of phantom cosmology (but without the
need for exotic forms of matter).

\end{abstract}


\

Cosmological data, including galaxy, CMB and supernovae
observations \cite{obs}, seem to be consistent with a cosmological
constant or a  dark phantom  energy with effective equation of
state parameter $\gamma<0$ \cite{phantom}. We shall study
cosmological models with a tilting, but otherwise conventional,
perfect fluid, and show that the models have dynamical behaviour
similar to that of phantom cosmology, but without the need for any
exotic forms of matter and consequently avoiding the pathologies
in these models, such as the existence of ghosts.

For spatially homogeneous (SH)  Bianchi cosmologies models, the
universe is foliated into space-like hypersurfaces \cite{DS}, and
there are two naturally defined time-like vectors: the unit vector
field, $n^{a}$, normal to the group orbits, and the four-velocity,
$u^{a}$, of the perfect fluid.  If $u^{a}$ is not aligned with
$n^{a}$, the model is called \emph{tilted} (and non-tilted or
orthogonal otherwise) \cite{KingEllis}. Usually, the kinematical
quantities associated with the normal congruence $n^a$ of the
spatial symmetry surfaces, rather than the fluid flow $u^a$, are
used as variables. Following \cite{KingEllis}, a tilt variable $v$
is introduced, so that in an orthonormal frame where
$n^a=(1,0,0,0)$, we have
\begin{equation}
  u^a=\frac{1}{\sqrt{1-v^2}}\left(1,v,0,0\right) .
\end{equation}

We will  assume a perfect-fluid matter source with
$p=(\gamma-1)\mu$ as equation of state, where $\mu$ is the energy
density, $p$ is the pressure, and $\gamma$ is a constant.
Causality then requires $\gamma$ to be in the interval
$0\leq\gamma\leq2$. A positive cosmological constant may also be
included in the models. Such SH tilted cosmologies with a
$\gamma$-law perfect fluid source have been studied by a number of
authors \cite{CollinsEllis,HWV,BHtilted,CH,HHC,HHLC,tilt}. It is
known that the tilt can become extreme ($v^2\rightarrow 1$) asymptotically to the
future. In particular, let $T$ be the proper time as measured
along the fluid congruence, and let us define the quantity $\Delta
T$:
\begin{equation}
\lb{def_T}
    \Delta T\equiv \int_{\tau_0}^{\infty}\frac 1H\sqrt{1-v^2}{\d
    \tau} 
\end{equation}
where $\tau$ is a dynamical time variable defined in terms of the
clock time $t$ and the Hubble scalar $H$ by $\frac{\d\tau}{\d
t}=H$. If $\Delta T$ is finite, then the fluid congruence is future
incomplete: the fluid observers will reach infinite expansion
within finite proper time. Therefore, in spite of the fact that
such a spacetime may be future geodesically complete
\cite{Rendall}, the worldlines defined by the fluid congruence
$u^a$  may become null with respect to the  normal congruence $n^a$, sometimes so quickly that this occurs within finite fluid proper time. 
We note that
for a SH cosmology for which $v^2\rightarrow 1$ (which is a
necessary condition for $\Delta T$ to be finite) the Hubble
parameter of the fluid congruence, $H_{\text{fluid}}$ also
diverges.


The locally rotational symmetric (LRS) Bianchi type V perfect fluid models were studied
in \cite{HWV}. The global structure of Bianchi type V  models was
studied in more detail in \cite{CollinsEllis}. It was shown that
for models with $4/3 < \gamma \leq 2$, to the future the tilt can
become extreme (in a finite time as measured along the fluid
congruence) and,  as $T$ approaches a finite limiting value, the
fluid  worldlines become null with respect to $n^a$, and its length scale $\ell$ and some of its kinematic variables (i.e., the expansion, the shear, the acceleration, and the vorticity) 
 diverge, but its matter density and
curvature scalars tend to zero. 
Although this peculiarity only occurs for ultra-radiative
fluids, which are unlikely to dominate the late universe, we will show that this behaviour can even occur for universes where ultra-radiative fluid is not dominant.

We shall first show that the fluid proper time is finite as the
solution approaches its asymptotic state (the extremely tilted Milne equilibrium point
$M^-$, defined by $ \Sigma=0$, $A=1$, $v=-1$, $\Omega =0$, $\Omega_\Lambda
=0$, and with deceleration parameter $q = 0$) for $4/3 <
\gamma < 2$ in the absence of a cosmological constant \cite{GE}.
The decay rates for the Hubble scalar $H=\theta/3$ and the
quantity $\sqrt{1-v^2}$ are $H \propto e^{-\tau}\,$,
 $\sqrt{1-v^2} \propto \exp\left[
    -(5\gamma-6)\tau/(2-\gamma) \right]$, whence
    \be
    \Delta T \propto \int_{\tau_0}^\infty \exp\left( \frac{-2(3\gamma-4)}{2-\gamma} \tau \right) \ \d\tau\,,
\ee
which is finite for $4/3 < \gamma < 2$.

Spatially homogeneous cosmological models with a positive
cosmological constant were investigated using dynamical systems
methods in \cite{GE}, extending the tilted LRS Bianchi type V analysis of
\cite{HWV} to the $\Lambda\neq0$ case. A de Sitter point with
extreme tilt is the future attractor for $\gamma > 4/3$.
Therefore, in general, for Bianchi type V models with
$\gamma>4/3$, the tilt again becomes extreme at late times and the
fluid motion is no longer orthogonal to the surfaces of
homogeneity.

Although it is known that expanding non-type-IX Bianchi models
with a positive cosmological constant isotropize to the future
(cosmic no hair theorem) \cite{Wald}, and that this result applies
to tilted models, the isotropization is with respect to the
congruence normal to the homogeneous symmetry surfaces -- not the
fluid congruence \cite{RaychaudhuriModak}. Thus in the Bianchi
type V models the space-time generically becomes de Sitter-like,
in accordance with the cosmic no hair theorem, but since the tilt
does not die away, isotropization of the cosmology does not occur
with respect to the fluid congruence.


This is a generic feature of spatially homogeneous models. In
general, SH models are not asymptotically isotropic. The spatially
homogeneous models do isotropize to the future in the presence of
a positive cosmological constant \cite{Wald}. By investigating the
asymptotic behaviour of a SH model with a cosmological constant
and a tilted perfect fluid with $p=(\gamma-1)\mu$ as the future de
Sitter model (with an extremely tilted perfect fluid) attractor is
approached we obtain $ H\propto H_0$ and: $ \Delta T\propto
\int_{\tau_0}^\infty \exp\left[- ({3\gamma-4})\tau/({2-\gamma})
\right] \ \d \tau\,$, which is finite for $4/3 < \gamma < 2$.
In the absence of a cosmological constant, in general SH models
are not asymptotically isotropic. Nevertheless, these models can spend
a long time close to a flat Friedmann model corresponding to a saddle
point. Considering one non-tilting perfect fluid with
$p_{\perp}=(\gamma_{\perp}-1)\mu_{\perp}$ and one tilting fluid
with $p=(\gamma-1)\mu$ (with $\gamma_{\perp}<\gamma$), as the
Friedmann equilibrium point is approached we obtain: $ \Delta
T\propto \int_{\tau_0}^\infty \exp\left[- ({3\gamma-4-\frac
32\gamma_{\perp}(2-\gamma)}) \tau/ ({2-\gamma} \right)] \ \d \tau.
$ This means that for solutions spending a finite (but arbitrarily
long) time close to the saddle $F$, exhibiting quasi-isotropic
behaviour consistent with observations, the Hubble parameter as
measured by the fluid, $H_{\text{fluid}}$, can become arbitrarily
large.

There will be different anisotropic asymptotic end-states
depending on Bianchi types. In the Bianchi type VII$_0$ model with
a tilted $\gamma$-law perfect fluid  \cite{HHLC,BarTip}, the future
asymptotic state for $\gamma>4/3$ was found to be anisotropic and
extremely tilted with $ H\propto e^{-2\tau}$ and $\Delta T
\propto
\int_{\tau_0}^{\infty}\exp\left[{(8-5\gamma)\tau/({2-\gamma}})\right]\d
\tau$. For models with $\gamma>8/5$, we find that this integral is
finite. 
The Bianchi type VIII
models are asymptotically extremely tilted for $1<\gamma<2$ and
the asymptotic solution is an extremely Weyl-curvature dominated
model \cite{BHWeyl,HLim}, with $H\propto \tau^{\frac
14}\exp\left(-\frac 32\tau\right)$, and $\Delta T\propto
\int_{\tau_0}^\infty
    \tau^{\frac{\gamma}{4(2-\gamma)}}\exp\left[-{3(3\gamma-4)}\tau/{2(2-\gamma)}\right]\d\tau$,
which is finite for $4/3<\gamma<2$.


It is known that for general spatially inhomogeneous perfect fluid
models with a cosmological constant, the de Sitter solution with
extreme tilt (where the tilt refers to the fluid tilt with respect
to a congruence with an acceleration that tends to zero) is
locally stable for $4/3 < \gamma < 2$ \cite{leuw,Rendall}.
From Equations (3.43) and (3.28) of \cite{leuw}, we have that $H
\propto H_0$ and 
\be \Delta T \propto \int_{\tau_0}^\infty
\exp\left( -\frac{3\gamma-4}{2-\gamma} \tau \right) \ \d \tau\, ,
\ee 
which is finite for $4/3 < \gamma < 2$, as required. We
note that for $4/3<\gamma<2$, the generic behaviour is
$v\rightarrow 1$, which implies that inflation does not isotropize
an ultra-radiative fluid. Even if inflation is turned off after a
certain number of $e$-foldings, $H_{\text{fluid}}$ can become
arbitrary large. Nevertheless, as noted above, this result does not
contradict the cosmic no-hair theorem \cite{Wald}.


Therefore, we have found that, for $\gamma > 4/3$, the fluid congruence becomes null with respect to the normal congruence in finite fluid proper time, and a 'kinematic singularity' develops for the fluid congruence.
 To fully understand the behaviour of these
models and their physical properties, the dynamics need to be
studied using a formulation adapted to the fluid (i.e., utilizing
a fluid-comoving frame). By using the boost formulae relating the normal and fluid congurences this singular behaviour for the fluid congruence can be confirmed.  

This mathematical instability might lead to some interesting
physics. Expanding universes that come to a violent end after a
finite proper time have arisen in a different context
\cite{phantom,Sudden}. Models with a constant equation of state parameter
$\gamma<0$, dubbed ``phantom energy'', lead to a singularity
commonly called the {\it big rip}. In this paradigm, during the
cosmic evolution the scale factor grows more rapidly than the
Hubble distance and consequently blows up in a finite proper time,
and is typically characterized by a divergent pressure and
acceleration. As the {\it big rip} singularity is approached, both
the strong and weak energy conditions are violated. For the so-called
\emph{sudden future singularities} \cite{Sudden} the strong energy
condition needs not be violated, yet a future singularity forms within
finite time. In our examples, no energy conditions are violated; in fact,
the energy density in the fluid frame, $\mu_{\rm fluid}$, tends to zero
as the singularity is approached.  



The Hubble scalar for the fluid frame can be computed from the boost
formula for $H$, and is of the form $H_{\rm fluid} = BH/\sqrt{1-v^2}$ \cite{long}.
The above examples lead to a diverging $H_{\rm fluid}$,
with $H_{\rm fluid} \propto e^{-(q_{\rm fluid}+1)\tauf}$,
and $\tauf = B \tau$ in the limit.
We then examine the value of the deceleration parameter $q_{\rm fluid}$
and compare with the critical value of $-1$. Equivalently, we can compare
with a phantom fluid in a flat, isotropic model using the effective
equation of state parameter $\geff$, given by 
\beq
\geff =\frac 23(1+q_{\rm fluid}),
\eeq 
with corresponding critical value of 0. 

In the general case of inhomogeneous cosmological models with a
cosmological constant, the de Sitter model with extreme tilt is
the future asymptotic state for $\gamma>4/3$, and we have  
\beq 
q_{\rm fluid}=-\frac 32(3\gamma-4)-1<-1.
\eeq
Consequently, as the de Sitter asymptotic state is approached
the dynamical effect of the ultra-radiative perfect fluid congurence behaves similarly to a
phantom energy in an isotropic and spatially flat spacetime. Note
that inflation does not stop the big rip from occurring.

Let us now briefly consider the SH models discussed above. For the
LRS Bianchi type V models, on the approach to the extremely tilted Milne solution (for $\gamma>4/3$) we find that 
\[ q_{\rm fluid} =-\frac 32(3\gamma-4)-1<-1. \]
For the Bianchi type VIII model with $\gamma>4/3$ we again obtain 
\[ q_{\rm fluid} =-\frac 32(3\gamma-4)-1<-1. \]
For the Bianchi type VII$_0$ model with $\gamma>8/5$ we obtain 
\[ q_{\rm fluid} =-\frac 32(5\gamma-8)-1<-1; \]
this difference arises due to the fact that the type VII$_0$ models are geometrically more special than Bianchi type VIII models.
Similarly, for the two-fluid example, we have 
\[ q_{\rm fluid}<-1,\] 
for $\gamma > \tfrac{2(4+3\gp)}{3(2+\gp)}$. A similar behaviour also occurs for Bianchi type VII$_h$ models (see \cite{long}). 
In the generic SH
models (such as, for example, the Bianchi type VIII model) the
ultra-radiative perfect fluid effectively behaves dynamically like a phantom
energy in the sense that the length scale and the Hubble scalar diverges as the future asymptotic state is approached. In the more
special examples, the requirement is that the threshold equation of state
must be equal to or higher than that of radiation.

Therefore, as the future asymptotic state is approached the
ultra-radiative perfect fluid effectively behaves like a phantom
energy in an isotropic and spatially flat spacetime. It is
important to note that the energy conditions of the perfect fluid
are nowhere violated.

Let us discuss the physical consequences of this dynamical
behaviour in a little more detail. 
One can consider models that spend a period close to isotropy (i.e.,
close to a Friedmann saddle point), with a small tilt. Thereafter, the
models begin to evolve away from isotropy. Since the tilt is non-zero,
for $\gamma>4/3$ the models generically evolve towards an asymptotic state
with extreme tilt. 
As the tilt instability sets in during the transient regime, observers moving with the tilting fluid will experience a transition from a decelerating expansion to an accelerating expansion, and later, extremely accelerating expansion mimicking that of a
phantom cosmology.  Moreover, unlike in conventional phantom cosmology, in
the models studied here there is no need for any exotic forms of
matter; conventional matter which is tilting suffices. In a
braneworld approach, accelerating universes can also result
without a cosmological constant or other form of dark energy
\cite{DGP}. Indeed, other pathologies, such as the existence of
ghosts, are avoided in the models described here. This is also the
case in alternative models to phantom cosmology which result from
alternative theories of gravity, theories with non-minimal
couplings, and models in which the dark energy and quintessence
field interact \cite{inter}.  In addition, as noted above, due to
the existence of future attractors with extreme tilt the dynamical
behaviour described here is generic.

Although the dynamics in the comoving dark energy models and the tilting
models are qualitatively similar, the actual quantitative (physical)
predictions of the two different models may differ (due to the different
physical transient time scales in each model).  For example, the age of the
universe as measured in the two models may differ.  As an example, let us
calculate numerically the age in a LRS Bianchi type V model which starts near
a flat FL model at the time of decoupling.  In a conventional comoving dust
model (with no dark energy) with present energy density $\Omega_{{\rm dust},0} \approx
0.2$, the maximum age is approximately 11.8 billion years (which, as is well
known, is on the low side compared with the ages of the oldest astrophysical
objects in the Universe).  In a comoving dust model with present
energy density $\Omega_{{\rm dust},0} \approx 0.2$ and a cosmological constant
$\Omega_{\Lambda,0} \approx 0.8$, the estimated age is about 15.0 billion years, which is
about a 27\% increase over the age for the dust model and consistent with
current observational data.  In a LRS Bianchi type V model with a comoving
dust model with $\Omega_{{\rm dust},0} \approx 0.2$ and a (second) tilting perfect fluid  with a total effective energy density
$\Omega_{{\rm tilt},0} \approx 0.8$, then according to the tilted fluid observer with present energy density $\Omega_{\rm dust,0,fluid ~frame} = 0.2$, the maximum age as measured in the tilting
fluid frame appears to be (depending on the initial conditions in the various
numerical experiments) about 13.2 billion years (this particular value occurs for the
initial conditions:  $A_0 = 0.641$, $\Sigma_{+,0} = 0.0056$, $v=-0.443$;
$\Omega_0 = 0.0111$).  This is a 12\% increase
over the original age of 11.8 billion years (and marginally consistent with
observations).  The situation is expected to be similar in the more general
Bianchi VIII models, although the quantitive predictions will depend on the
precise initial conditions and it is possible more fine tuning may be necessary.

Therefore, although qualitatively the calculated age of the universe in both the dark
energy models and the tilting fluid models are similar in that the age is
greater than in the conventional models (without dark energy or a second
tilting fluid), quantitatively the increased age appears to be greater in the
dark energy models.  However, the tilted fluid models are still 
physically viable.  
In future work we shall further study physical predictions of the  tilting fluid
models. In addition to the age problem, it is also 
of interest to investigate the effect of
tilt on cosmic microwave background radiation observations and whether
these models offer a possible explanation for
the various anomalies on large angular
scales found in the WMAP data \cite{Jaffe,Oliveira-Costa}.

\section*{Acknowledgment}
This work was supported by a Killam Postdoctoral Fellowship (SH)
and the Natural Sciences and Engineering Research Council of
Canada (AC).

\end{spacing}


\begin{thebibliography}{99}



\bibitem{obs}
A.G. Riess et al., \textit{Ap. J.} {\bf 607} (2004) 665; U.
Seljak et al., \textit{Phys.  Rev.  D} {\bf 71} (2005) 103515.

\bibitem{phantom}
R.R.~Caldwell, \textit{Phys. Lett. B} {\bf 545} (2002)  23; B.
McInnes, {\sl JHEP} {\bf 0208} (2002) 029; R.R.~Caldwell, M.~
Kamionkowski, N.N.~Weinberg, \textit{Phys.  Rev.  Lett.} {\bf 91}
(2003) 071301; L.P. Chimento and R. Lazkoz, \textit{Phys.  Rev.
Lett.}  {\bf 91} (2003) 211301; A. Coley, S. Hervik and J. Latta,
[astro-ph/0503169].



\bibitem{DS} G.F.R. Ellis and M.A.H. MacCallum, \textit{Comm. Math. Phys.}
\textbf{12} (1969) 108;
J. Wainwright and G.F.R. Ellis, eds, \textit{Dynamical Systems in Cosmology}, Cambridge University Press (1997);
A.A. Coley, \textit{Dynamical Systems and Cosmology},
Kluwer, Academic Publishers (2003);
J.D. Barrow and D.H. Sonoda,
\textit{Phys. Reports} \textbf{139} (1986) 1.


\bibitem{KingEllis} A.R. King and G.F.R. Ellis, \textit{Comm. Math. Phys.}
\textbf{31} (1973) 209.

\bibitem{tilt} C.G. Hewitt, R. Bridson, J. Wainwright, \textit{Gen. Rel. Grav.%
} \textbf{33} (2001) 65;
I.S. Shikin, \textit{Sov. Phys. JETP} \textbf{41} (1976) 794;
C.B. Collins, \textit{Comm. Math. Phys.} \textbf{39} (1974) 131;
S. Hervik, \textit{Class. Quantum Grav.} \textbf{21} (2004) 2301;
A.A. Coley and S. Hervik,  \textit{Class. Quant. Grav.} \textbf{21} (2004) 4193.

\bibitem{CollinsEllis} C.B. Collins and  G.F.R. Ellis, \textit{Phys. Rep.}
{\bf 56} (1979) 65.


\bibitem{HWV} C.G. Hewitt and J. Wainwright, \textit{Phys. Rev.} \textbf{D46}
(1992) 4242.


\bibitem{BHtilted} J.D. Barrow and S. Hervik, \textit{Class. Quant.
  Grav.} \textbf{20} (2003) 2841.

\bibitem{CH} A.A. Coley and S. Hervik,  \textit{Class. Quant. Grav.} \textbf{22} (2005) 579.

\bibitem{HHC} S. Hervik, R.J. van den Hoogen and A.A. Coley, \textit{Class. Quant. Grav.} \textbf{22} (2005)
607.

\bibitem{HHLC}  S. Hervik, R.J. van den Hoogen,
W.C. Lim and A.A. Coley, \textit{Class. Quant. Grav.} \textbf{23}
(2006) 845.



\bibitem{Rendall}
A.D. Rendall, \textit{Math. Proc. Camb. Phil. Soc.} \textbf{118}
(1995) 511.


\bibitem{GE} M. Goliath and G.F.R. Ellis, \textit{Phys. Rev. D} {\bf 60} (1999) 023502.

\bibitem{Wald} R.M. Wald, \textit{Phys. Rev. } \textbf{D28} (1983)
2118.

\bibitem{RaychaudhuriModak}
A.K. Raychaudhuri and B. Modak, \textit{Class. Quant. Grav.} {\bf
5} (1988) 225.


\bibitem{BarTip} J.D. Barrow and F.J. Tipler, \textit{Nature} \textbf{276} (1978) 453.

\bibitem{BHWeyl}
J.D. Barrow and S. Hervik, \textit{Class. Quant. Grav.}
\textbf{19} (2002) 5173.

\bibitem{HLim}S. Hervik and W.C. Lim, \textit{Class. Quant. Grav.} \textbf{23} (2006) 3017.



\bibitem{leuw}
W.C. Lim, H. van Elst, C. Uggla and J. Wainwright, \textit{Phys.
Rev. D} {\bf 69} (2004) 103507.

\bibitem{Sudden}
J.D. Barrow, \textit{Class. Quant. Grav.} \textbf{21} (2004) L79; 
 J.D. Barrow and C.G. Tsagas,  \textit{Class. Quant. Grav.} \textbf{22} (2005) 1563; 

\bibitem{long}
A.A. Coley, S. Hervik, W.C. Lim, \textit{Class. Quant. Grav.}
\textbf{23} (2006) 3573. 

\bibitem{DGP}
G. Dvali, G. Gabadadze and M. Porrati \textit{Phys. Lett. B} {\bf 485} (2000)  208;
V. Sahni and V.Yu. Shtanov, \textit{JCAP} {\bf 11} (2003)  014.

\bibitem{inter}
L. Amendola, \textit{Phys.  Rev.  D} {\bf 62} (2000) 043511;
G. Huey and B.D. Wandelt, [astro-ph/0407196];
S. Das, P.S. Corasanuti and J. Khoury, [astro-ph/0510628].


\bibitem{Jaffe} T. R. Jaffe, A. J. Banday, H. K. Eriksen, K. M. Gorski and F. K.
Hansen,  \textit{Astrophys. J.} {\bf 629} (2005) L1
[astro-ph/0503213]; T. R. Jaffe, S. Hervik, A. J. Banday and K. M.
Gorski, to appear in \textit{Astrophys. J.}  [astro-ph/0512433].


\bibitem{Oliveira-Costa} A. de Oliveira-Costa, M. Tegmark, M. Zaldarriaga
and A. Hamilton, \textit{Phys. Rev. D} {\bf 69} (2004) 063516
[astro-ph/0307282].



\end{thebibliography}
\end{document}